\documentclass{emulateapj}
\RequirePackage{lineno}
\usepackage{amssymb}
\usepackage{amsmath}
\usepackage{mathrsfs}
\usepackage{natbib}
\usepackage[colorlinks, linkcolor=blue, urlcolor=blue, citecolor=blue]{hyperref}
\usepackage{array}
\usepackage{float}
\usepackage{graphicx}
\usepackage{subfigure}
\usepackage{color}
\usepackage[all]{hypcap}
\usepackage[toc,page]{appendix}

\def\beq{\begin{equation}}
\def\enq{\end{equation}}
\def\bea{\begin{eqnarray}}
\def\ena{\end{eqnarray}}

\begin{document}

\title{Prompt Neutrino Emission of Gamma-Ray Bursts in the Dissipative Photospheric Scenario Revisited: Possible Contributions from  Cocoons}
\author{Di Xiao\altaffilmark{1,2}, Zi-Gao Dai\altaffilmark{1,2}, and Peter M\'esz\'aros\altaffilmark{3}}
\affil{\altaffilmark{1}School of Astronomy and Space Science, Nanjing University, Nanjing 210093, China; dzg@nju.edu.cn}
\affil{\altaffilmark{2}Key Laboratory of Modern Astronomy and Astrophysics (Nanjing University), Ministry of Education, China}
\affil{\altaffilmark{3}Center for Particle and Gravitational Astrophysics; Department of Physics; Department of Astronomy and Astrophysics, The Pennsylvania State University, University Park, PA 16802, USA}


\begin{abstract}
High-energy neutrinos are expected to originate from different stages in a gamma-ray burst (GRB) event. In this work we revisit the dissipative photospheric scenario, in which the GRB prompt emission is produced around the photospheric radius. Meanwhile, possible dissipation mechanisms (e.g., internal shocks or magnetic reconnection) could accelerate cosmic-rays (CRs) to ultra-high energies and then produce neutrinos via hadronuclear and photohadronic processes, which are referred to as prompt neutrinos. In this paper, we obtain the prompt neutrino spectrum of a single GRB within a self-consistent analytical framework, in which the jet-cocoon structure and possible collimation effects are included. We investigate a possible neutrino signal from the cocoon, which has been ignored in the previous studies. We show that if a GRB event happens at a distance of the order of Mpc, there is a great chance to observe the neutrino emission from the cocoon by IceCube, which is even more promising than jet neutrinos since the opening angle of the cocoon is much larger. We also figure out the diffuse neutrino flux of GRB cocoons and find that it could be comparable with that of the jets. Our results are consistent with the latest result reported by the IceCube collaboration that no significant correlation between neutrino events and observed GRBs is seen in the new data \citep{aar17}.
\end{abstract}


\section{Introduction}
Relativistic jets are universal in long-duration gamma-ray burst (GRB) models. Before breaking out, a relativistic jet must propagate in the progenitor envelope, leading to the formation of a cocoon structure which in turn collimates the jet \citep{mat03, mor07, miz09, bro11, miz13}. In this stage, a part of the jet energy is injected into the cocoon. After breakout, as suggested by the photospheric scenario \citep[e.g.][]{ree05, mur08ph, wan09, gao12}, a significant fraction of the outflow energy is converted to the radiation energy via some dissipation mechanism within the outflow (e.g., internal shocks or magnetic reconnection) and the jet produces prompt gamma-ray emission around the photospheric radius \citep{mur08ph}. At the same time, the cocoon also expands adiabatically like a fireball \citep{miz13}. Since the amount of energy in the cocoon is much smaller than the GRB isotropic equivalent energy, the cocoon emission is always ignored in the previous works. Recently, \citet{nak17} studied the observable signatures of GRB cocoons in detail, which also motivates us to investigate the neutrino emission from the cocoon.

GRBs have been discussed as capable cosmic-ray (CR) accelerators and are possible sources for high energy neutrinos \citep[e.g.][]{wax97, wax99, rac98, alv00, bah01, gue03, mura06, bec08}. We can expect high-energy neutrinos originating from different stages in a GRB event. First, neutrinos could be produced while the jet is still propagating in the envelope \citep[e.g.][]{mes01, hor08, mi13, xia14}. However, the forward and reverse shock inside the progenitor may be radiation-mediated \citep{lev08, kat10}, and therefore these neutrinos are more relevant for low-luminosity GRBs or chocked jets \citep{liu11, mi13, xia14, xia15, sen16}. They appear as a precursor burst and their energies are mainly sub-PeV \citep{xia14}. Second, the prompt photon emission from GRBs can be correlated to the production of neutrinos, since protons are also believed to be accelerated in relativistic internal shocks \citep[e.g.][]{vie95, wax97, wax99, mes00, gue04, mur06, liu13}. Alternatively, the neutrino emission might arise from a dissipative jet photosphere \citep[e.g.][]{ree05, mur08ph, wan09, gao12, zha13}. Finally, there will be neutrino production in the afterglow phase since external forward and reverse shocks could also accelerate CRs \citep[e.g.][]{wax00, dai01, der03, raz13, wan15}.

In this work, we focus on the neutrino production in the prompt emission phase, which are referred to as ``prompt neutrinos''. We carry on our calculations within the dissipative photospheric scenario, in which the dissipation and thermalization should occur around or below the photosphere \citep{mur08ph, wan09}. Most importantly, we try to figure out the neutrino emission from the cocoon, which is largely ignored in previous works. We show that there is a great chance to observe the cocoon neutrinos if the GRB is located at a distance of the order of Mpc. If the observer is off-axis, the only neutrino signal from a GRB source we can observe originates from the cocoon. These cocoon neutrinos would be falsely identified as not correlated with a GRB, which is consistent with the null results of searching for GRB neutrinos by IceCube \citep[e.g.][]{aar12, aar17}.

The paper is organized as follows. We introduce the model and dynamics in Section 2. Then we consider the prompt neutrino emission of a nearby individual GRB in Section 3. Section 4 presents our results of the diffuse prompt neutrino flux from GRBs, compared with the
IceCube data. Finally, we provide a summary and discussion in Section 5.

\section{Model and Dynamics}
First, we consider the neutrino emission of a typical GRB. We assume that the jet luminosity is $L_j=10^{50}\,\rm erg\,s^{-1}$ with an initial opening angle $\theta_0=0.1$, and the progenitor envelope has a power-law density profile $\rho_a(r)=Ar^{-2}$, where $A=M_\ast/4\pi R_\ast=1.59\times10^{22}\,\rm g\,cm^{-1} $ for the typical progenitor mass $M_{\ast}=10M_\odot$ and radius $R_\ast=10^{11}\,\rm cm$. According to \citet{bro11}, we can define the dimensionless $\tilde{L}\equiv L_j/(\Sigma_j\rho_ac^3)$, where $\Sigma_j$ is the jet's cross section. For our parameters taken here, we get $\tilde{L}\simeq 2.7$ and it falls into $1\leq\tilde{L}<\theta_0^{-4/3}$ regime \citep{bro11, xia14}, which means that the jet is dramatically collimated by the cocoon. Therefore, the head velocity is determined by $\beta_h=\beta_0/(1+\tilde{L}^{-1/2})\simeq0.62$, where the jet material velocity is relativistic $\beta_0\simeq1$ (inside the star, $\Gamma_0\sim\theta_0^{-1}\sim10 $ \citep{miz13}). The time for jet breakout is $t_b=R_\ast/(\beta_hc)\simeq5.4\,\rm s$ so that the energy flowed into the cocoon is $E_c=L_jt_b(1-\beta_h)\simeq2.0\times10^{50}\,\rm erg$. This is just a conservative estimate of cocoon energy since there might be extra energy injection into the cocoon in the afterglow phase, which we will discuss later. The density of the jet material is $\rho_j(r)=L_j/(\Gamma_0^2\pi r^2\bar{\theta}_0^2c^3)$, where $\bar{\theta}_0 < \theta_0$ is the opening angle after collimation. During the jet propagating in the envelope, the Thomson scattering depth $\tau_T=(\rho_j/m_p)\sigma_T(R_\ast/\Gamma_0)\gg 1$, so that both the forward shock and reverse shock formed at the jet head are radiation-mediated \citep{mi13}. Efficient Fermi acceleration would not be achieved inside the star and we can ignore the high energy neutrino emission when the jet is still propagating in the envelope.

At the time of breakout, the jet expands to an opening angle $\theta_j\sim1/\Gamma_0\sim\theta_0$ \citep{mor07, laz09, bro11, laz13}. The shape of the cocoon is roughly a barrel with a height $R_\ast$ and a width $\sim R_\ast\theta_j$ \citep{bro11, nak17}. The total cocoon volume is $V_c\simeq\pi R_\ast^3\theta_j^2$ and the blackbody temperature is $T_c=\left[E_c/(aV_c)\right]^{1/4}$, where the radiation constant $a=7.5657\times10^{-15}\,\rm erg\,cm^{-3}\,K^{-4}$. After breakout, a fraction of the jet's internal energy is converted to kinetic energy, and the typical Lorentz factor of the jet material is $\Gamma_j\sim 10^2$.  The photospheric radius of the jet is determined by $\tau_T(r)=1$, which gives
\beq
R_{\mathrm{ph},j}=L_j\sigma_T/(\Gamma_j^3\pi\theta_j^2m_pc^3)\simeq1.5\times10^{12}\,\rm cm.
\enq
At the same time, the cocoon expands adiabatically like a fireball and generally its photospheric radius is larger than the jet, which we will show later.

\section{Prompt Neutrino Emission of a Single GRB}
\subsection{Jet neutrinos}
We assume that the dissipation mechanism is an internal shock. Let us consider an internal collision of two subshells, with different Lorentz factors $\Gamma_s=10^2$ and $\Gamma_f=10^3$. Hence, the Lorentz factor of the internal shock in their center-of-mass frame is $\Gamma_{\rm rel}\sim[(\Gamma_f/\Gamma_s+\Gamma_s/\Gamma_f)/2]^{1/2}\sim2.2$, which is mildly relativistic and capable of accelerating CRs. At $R_{\mathrm{ph},j}$, the postshock density is $\rho_j^{\prime}=(4\Gamma_{\rm rel}+3)\rho_j(R_{\mathrm{ph},j})$, and the internal energy is $e_{\rm int}^{\prime}=(4\Gamma_{\rm rel}+3)(\Gamma_{\rm rel}-1)\rho_j(R_{\mathrm{ph},j})c^2$, and thus the equipartition magnetic field is $B^{\prime}=(8\pi\epsilon_Be_{\rm int}^{\prime})^{1/2}$, the temperature is $kT^{\prime}=\{[15(hc)^3/(8\pi^5)]\epsilon_ee_{\rm int}^{\prime}\}^{1/4}$, where $\epsilon_B=\epsilon_e=0.1$ is assumed, and the number density of thermal photons is $n_\gamma^{\prime}=19.232\pi\times(kT^{\prime})^3/(hc)^3$.

High-energy protons lose their energies through adiabatic, radiative and hadronic processes. The adiabatic cooling timescale $t_{\rm ad}$ is comparable to the dynamical timescale. Radiative cooling includes synchrotron and inverse Compton (IC) scattering and hadronic cooling mechanisms mainly contain inelastic $pp$ collisions, the Bethe-Heitler pair production and photomeson production processes. In this work, we calculate the cooling rates of high-energy protons using the numerical method described in \citet{mur07,mur08ph} and \citet{xia16b}. We plot in Figure 1 the proton cooling rates at the jet's photospheric radius, also shown is the Fermi acceleration timescale (red solid line), which is $t_{\rm acc}=\eta\epsilon_p/(eBc)$ with $\eta\sim1-10$ \citep{raz04, ando05, mur08ph}. The maximum proton energy can be found by equaling the total energy loss rate to the acceleration rate.

\begin{figure}
\plotone{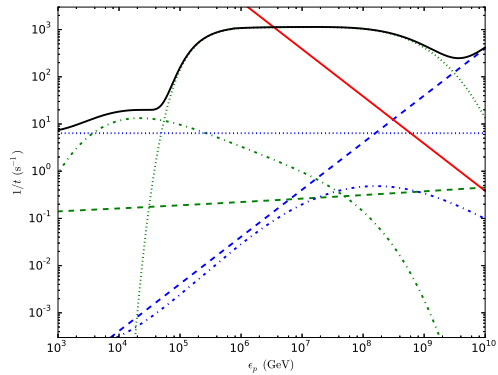}
\caption{The inverse of cooling timescales for protons at the jet's photospheric radius: blue dashed line -- synchrotron, blue dotdashed -- IC, blue dotted line -- adiabatic cooling, green dashed line -- inelastic $pp$ scattering, green dot-dashed line -- Bethe-Heitler process, green dotted line -- photomeson production, black solid line -- total. Also shown is the Fermi acceleration timescale -- red solid line.
\label{fig1}}
\end{figure}

We assume the initial accelerated proton spectrum to be a power-law $dN_p/d\epsilon_p\propto\epsilon_p^{-s}$ with index $s=2$, and we consider two suppression factors which modify the neutrino spectrum. The suppression factor due to proton cooling can be written as $\zeta_{\rm CRsup}$ \citep{mur08ph,wan09,xia16b},
\beq
\zeta_{\rm CRsup}(\epsilon_\nu)=\frac{t_{pp}^{-1}+t_{p\gamma}^{-1}}{t_{\rm syn}^{-1}+t_{\rm IC}^{-1}+t_{\rm ad}^{-1}+t_{pp}^{-1}+t_{\rm BH}^{-1}+t_{p\gamma}^{-1}},
\label{eq:suppr}
\enq
where $t_{\rm syn}$, $t_{\rm IC}$, $t_{\rm ad}$, $t_{pp}$, $t_{\rm BH}$, and $t_{p\gamma}$ represent synchorotron, inverse Compton, adiabatic, inelastic $pp$ collision, the Bethe-Heitler and $p\gamma$ photomeson production timescales for protons, respectively.
Furthermore, the suppression factor due to meson cooling is expressed by \citep{xia16b}
\beq
\zeta_{\pi \rm sup}(\epsilon_\nu)=\frac{t_{\rm dec}^{-1}}{t_{\rm dec}^{-1}+t_{\rm syn}^{-1}+t_{\rm had}^{-1}},
\label{eq:pionsup}
\enq
where $t_{\rm dec}$, $t_{\rm rad}$, and $t_{\rm had}$ are the decay, radiative loss and hadronic loss timescales for pions, respectively.

The jet's neutrino fluence of a single nearby GRB can be expressed as
\bea
E_\nu^2\mathcal{F}_{\nu_i}(E_\nu)=\frac{K}{4(1+K)}&\times&
\frac{\epsilon_{\rm acc}E_{\rm iso}}{4\pi D_L^2\ln{(E_{p,\rm max}/E_{p,\rm min})}}\nonumber\\
&\times&\zeta_{\rm CRsup}(E_\nu)\zeta_{\pi \rm sup}(E_\nu),
\label{eq:fluence}
\ena
where $K$ denotes the average ratio of charged to neutral pions, with $K\approx1$ for $p\gamma$ and
$K\approx2$ for $pp$ interactions \citep{mur16a}. In this work we just assume that the CR acceleration efficiency $\epsilon_{\rm acc}=0.1$. In Figure 2 we plot the jet's neutrino fluence, assuming that the
GRB has $E_{\rm iso}= (4L_j/\theta_j^2)T_{\rm dur}=4\times10^{53}\,\rm erg$ with a typical jet duration of $10\,\rm s$ and is at a distance $D_L=100\,\rm Mpc$ with a jet pointing to us.

\begin{figure}
\begin{center}
\plotone{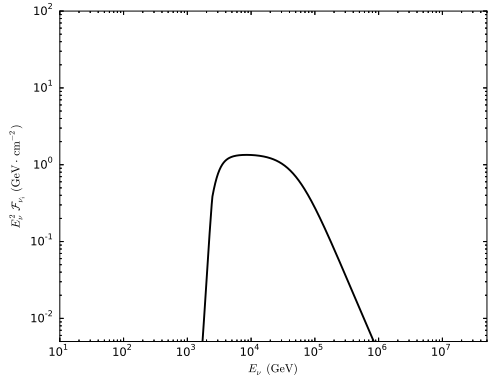}
\caption{The neutrino spectrum (per flavor) of a single GRB jet. The flux drop on high-energy end is due to meson cooling.
\label{fig2}}
\end{center}
\end{figure}

The neutrino spectrum of the jet is characterized by a global suppression which is caused by the other cooling mechanisms competing with $pp$ and $p\gamma$ neutrino production processes, and a sharp drop at energies $>100$ TeV due to the radiative cooling suppression of pions. The low-energy neutrino flux is very faint since there is an threshold energy for the $p\gamma$ interaction and the $pp$ efficiency $f_{pp}=t_{\rm dyn}/t_{pp}\ll1$ at the photospheric radius. The fluence is so high that it should be detected by IceCube if the observer is on-axis. However, the chance is very small since the jet is highly beamed with an opening angle $\sim0.1$. On average, one of every $\sim f_b^{-1}\sim 400$ GRBs has a jet pointing to us, where $f_b\equiv(1-\cos\theta_j)/2\simeq \theta_j^2/4$ is the beaming factor.

\subsection{Cocoon neutrinos}
Once the jet breaks out, the cocoon will also break out and expand as a fireball, with a total energy $E_c$. The cocoon could be divided into two parts: isotropic shocked stellar materials and shocked jet materials with typical opening angle $\theta_c\sim0.5$ \citep{bro11, nak17}. Numerical results show that there is also a distribution of Lorentz factors at different positions of shocked jet materials \citep{nak17}, ranging from non-relativistic $\Gamma_{c,s}=1$ to relativistic $\Gamma_{c,f}=10$. Hence, in the case of ``partial mixing" \citep{nak17}, we also expect that two subshells collide with each other and then resultant shocks can accelerate CRs. Similar to that in the jet, we have $\Gamma_{\mathrm{rel}, c}\sim2.2$.

The photospheric radius of the cocoon is
\beq
R_{\mathrm{ph},c}=\left(\frac{\kappa E_cf_{\Gamma_c}}{2\pi\theta_c^2 c^2\Gamma_c}\right)^{1/2},
\label{eq:corph}
\enq
where $\kappa=0.2\,\rm cm^2\,g^{-1}$ is the mean opacity and $f_{\Gamma_c}$ represents the fraction of energy carried by materials with Lorentz factor $\Gamma_c$. Numerical results suggest that the cocoon energy is divided uniformly for every logarithmic scale of $\Gamma_c\beta_c$ in the range of $0.1-10$ and then $f_{\Gamma_c}\sim 0.1$ \citep{nak17}. At $R_{\mathrm{ph},c}$, the volume of the shell in the comoving frame is $V_{\rm ph}=2\pi\theta_c^2R_{\mathrm{ph},c}^3/\Gamma_c$. Since the cocoon mass is $M_c=f_{\Gamma_c}E_c/\Gamma_cc^2$, the preshock density is then $\rho_c=M_c/V_{\rm ph}$. After adiabatic expansion to the photospheric radius, the cocoon temperature cools to $T_c^{\prime}=T_c(V_cf_{\Gamma_c}/V_{\rm ph})^{1/3}$. The postshock internal energy can be expressed as $e_{\mathrm{int},c}^{\prime}=(4\Gamma_{\mathrm{rel},c}+3)(\Gamma_{\mathrm{rel},c}-1)\rho_cc^2$, and then the remaining calculation is similar to that for the jet. Adopting a value of $E_c=2\times10^{50}\,\rm erg$ as in Section 2, we plot the cooling rates of protons at the
cocoon's photospheric radius in Figure 3 and the spectrum of cocoon neutrinos in Figure 4.

\begin{figure}
\plotone{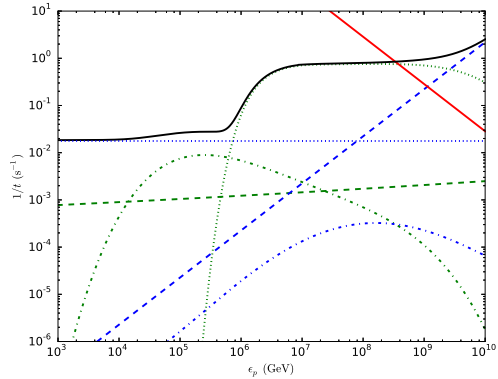}
\caption{Same as Figure 1, but at the cocoon's photospheric radius.
\label{fig3}}
\end{figure}

The spectrum of cocoon neutrinos peaks at a higher energy compared to jet neutrinos, since the
maximum CR energy can be much higher in the cocoon. The physical reason for this result
is that the photospheric radius of the cocoon is much larger than the jet according to equations (1) and (5), which leads to lower internal energy and blackbody temperature thus less proton cooling. The spectrum of such a cocoon is shown in Figure 4 for a distance $D_L=100\,\rm Mpc$.
The total neutrino fluence of this cocoon is two or three orders of magnitude fainter than
that of the jet neutrinos, because the total isotropic energy of the cocoon is much lower.

\begin{figure}
\begin{center}
\plotone{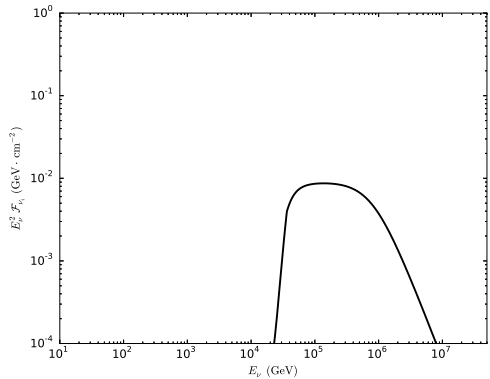}
\caption{The neutrino spectrum (per flavor) of a single GRB cocoon. Generally, the neutrino fluence from the cocoon peaks at higher energies, but is two or three magnitude fainter than that from the jet under conservative estimate.
\label{fig4}}
\end{center}
\end{figure}

Using the latest IceCube effective area $A(E_\nu)$ given by \citet{aar15c}, we can now estimate
the number of neutrino events in the IceCube detector. The number of muon neutrinos above 1~TeV is
\beq
N(>1\mathrm{TeV})=\int_{1\mathrm{TeV}}^{E_{\nu,\max}}{dE_\nu A(E_\nu)\mathcal{F}_\nu(E_\nu)}.
\label{eq:num}
\enq
For our parameters, the number of neutrinos from the jet is $N_j(>1\mathrm{TeV})\sim20$
for a typical jet duration of $10\,\rm s$ and the number of neutrinos from the cocoon
is $N_c(>1\mathrm{TeV})\sim0.1$. Note that the fluence depends on the inverse of the distance
square. Therefore, for a possible IceCube observation of the jet neutrinos, this source should be within
$D_{j,L}^\prime=(20/1)^{1/2}\times100\,\rm Mpc\sim440\,Mpc$ from the earth. This implies that on average
we would have to wait for $[(4/3)\pi D_{j,L}^{\prime3}\times\rho(0)]^{-1}\sim 3\,\rm yr$, where we assume the local burst rate $\rho(0)\sim1\,\rm Gpc^{-3}\,yr^{-1}$ \citep{gue05, pel08}. This time is inversely proportional to the jet duration. For the cocoon neutrinos, $D_{c,L}^\prime=(0.1/1)^{1/2}\times100\,\rm Mpc\sim32\,Mpc$ and the average waiting time is $[(4/3)\pi D_{c,L}^{\prime3}\times(\theta_c/\theta_j)^2\times\rho(0)]^{-1}\sim 290\,\rm yr$, where the second factor in the the square brackets has been introduced due to the effect of a wider outflow of the cocoon.

Since the nearby GRB rate is dominated by low-luminosity GRBs (LL-GRBs), it is of interest to investigate whether we can observe the cocoon neutrinos from them. To an order of magnitude estimate, a typical LL-GRB has a jet luminosity $L_{\rm j, LL}\sim10^{45}-10^{48}\,\rm erg\,s^{-1}$ \citep[e.g.][]{bro11b}. For a nominal value $10^{46}\,\rm erg\,s^{-1}$, the energy injected into the cocoon is at most $E_{\rm c,LL}=L_{\rm j,LL}T_{\rm dur}\sim10^{47} \,\rm erg$ (in the case of a chocked jet). This energy is three orders of magnitude lower than that of a normal GRB cocoon. We calculate the neutrino fluence of this nominal LL-GRB as before and find that the number of neutrinos in IceCube detector is $N_{\rm c, LL}=1.7\times10^{-4}$ if it is located at $100\,\rm Mpc$, leading to $D_{\rm LL}^{\prime}=1.3\,\rm Mpc$.  Although the local burst rate is much higher for LL-GRBs, $\rho_{\rm LL}(0)\sim200\,\rm Gpc^{-3}\,yr^{-1}$ \citep[e.g.][]{sun15}, assuming that the cocoon still expands to an opening angle $\theta_{\rm c, LL}\sim0.5$, the average waiting time is about $[(4/3)\pi D_{\rm LL}^{\prime3}\times(\theta_{\rm c,LL}/\theta_{\rm j,LL})^2\times\rho_{\rm LL}(0)]^{-1}\sim2\times10^4\,\rm yr$. Therefore, it may be not easy to detect neutrinos from LL-GRB cocoons.

Note that the cocoon fluence of the black solid line in Figure 4 is a rather conservative
estimate, since we have assumed that there is no extra energy deposited into the cocoon after
breakout. However, a large fraction of long GRBs have an X-ray afterglow plateau which lasts
$\sim 500\,\rm s$ to $10^3\,\rm s$ with a luminosity of $10^{46-50}\,\rm erg\,s^{-1}$
\citep[e.g.][]{dain15}. The total energy in the plateau may be comparable or greater than that
of the jet. One possible explanation for this plateau is that there is continued injection
of energy at a lower bulk Lorentz factor \citep[e.g.][]{zha06}, which would lead to continuous addition of
energy to the cocoon for $\sim 500-1000$ seconds.
The shock that causes the X-ray plateau would also accelerate CRs and make neutrinos.
Thus, the neutrino fluence could be higher by a factor of $\sim100$ and the waiting time
would be shorter by the same factor. Of course, Poisson fluctuations or a higher rate or efficiency could conceivably also reduce
this waiting time.

For individual source detections,
since the opening angle is as large as $\theta_c\sim0.5$, there is a great chance to observe
the neutrino emission from the cocoon if the GRB is located at a distance of the order of Mpc in the impulsive jet/cocoon model (or $\sim 100 \,\rm Mpc$, in the continuous injection plateau
model), although this still remains a difficult observation. With the upgrade process to IceCube-Gen2 in next few years, the effective area is larger, thus offering better hopes for detecting such low fluences of the cocoons.
Cocoon neutrinos are produced at $(R_{\rm ph,c}-R_{\rm ph,j})/c\sim 500\,\rm s$ later than the
jet neutrinos. For an on-axis observer, we can observe two neutrino signals. Jet neutrinos
come first with a larger fluence, followed by cocoon neutrinos. If the observing angle falls
into the range $\theta_j<\theta_{\rm obs}<\theta_c$, the jet neutrinos are missing and the
cocoon neutrinos are observable. However, for too large observing angle $\theta_{\rm obs}>\theta_c$,
no prompt neutrinos are expected. IceCube has reported no significant direct correlation between
the existing high energy starting event (HESE) with a GRB \citep{aar12,aar16,aar17}.
However, this constraint is not applicable for cocoon neutrinos.

\section{Diffuse Neutrino Flux of GRBs}
\label{sec:difnuflux}
To calculate the diffuse neutrino flux from GRBs, we have to assume the burst rate and luminosity function of GRBs. Here we adopt the best-fitting rate of the {\it Swift}-observed long GRBs \citep{wand10},
\beq
\rho(z)\propto\begin{cases}
(1+z)^{n_1} & z<z_1, \\
(1+z_1)^{n_1-n_2}(1+z)^{n_2} & z>z_1,
\end{cases}
\enq
where $z_1=3.1_{-0.6}^{+0.8},\,n_1=2.1_{-0.6}^{+0.5}$, and $n_2=-1.4_{-1.0}^{+2.4}$. The luminosity function is
\beq
\Psi(L)\equiv \frac{dN}{dL}\propto\begin{cases}
(L/L_b)^{-\alpha} & L<L_b,\\
(L/L_b)^{-\beta} & L>L_b,
\end{cases}
\enq
where $L_b=10^{52.5\pm0.2}\,{\rm erg}\,{\rm s}^{-1}$ is the break point of the GRB isotropic luminosity and the best-fitting indices are $\alpha=1.2_{-0.1}^{+0.2}$ and $\beta=2.4_{-0.6}^{+0.3}$ \citep{wand10}\footnote{Please note that our
definition of $\alpha$ and $\beta$ is slightly different from \citet{wand10}, where $dN/d\log L$ is used.}.
For completeness, we discuss the results based on different rates and luminosity functions of GRBs later in Section 5.

The diffuse prompt neutrino flux of GRBs is expressed as
\bea
E_{\nu}^2\Phi_{\nu_i}&=&\frac{c}{4\pi H_0}\times\frac{K}{4(1+K)}\nonumber\\
&\times&\int_0^{z_{\rm max}}dz\int_{L_{\min}}^{L_{\max}}dL\frac{1}{(1+z)^2\sqrt{\Omega_M(1+z)^3+\Omega_\Lambda}} \nonumber\\
&\times&\frac{\epsilon_{\rm acc}\rho(z)\Psi(L)E_{\rm iso}(L)\zeta_{\rm CRsup}(\epsilon_\nu,L)\zeta_{\rm \pi sup}(\epsilon_\nu,L)}{\ln \left(\epsilon_{p,\max}(L)/\epsilon_{p,\min}(L)\right)},\nonumber\\
\label{eq:diffnu}
\ena
where we adopt $L_{\min}=10^{45}\,\mathrm {erg\,s^{-1}},\,
L_{\max}=10^{52}\,\mathrm {erg\,s^{-1}},\,z_{\max}=20$ and the cosmological parameters
are $H_0=67.8\,{\rm km\,s^{-1}\,Mpc^{-1}}$ and $\Omega_M=0.308$ \citep{pla15}.

Figure 5 shows the diffuse neutrino flux from GRBs, compared with the observation of IceCube (blue datapoints, \citet{aar15}). The blue solid line corresponds to the jet neutrinos and the black solid line to cocoon neutrinos under conservative estimate. The Waxman-Bahcall bound is indicated by the green dashed lines. The prompt neutrino flux of GRB jets is consistent with the new constraints given by the IceCube Collaboration for the photospheric model \citep{aar17}. The uncertainty in total cocoon energy result in an range of the final diffuse flux of cocoon neutrinos, which is shown by the cyan region in Figure 5. The diffuse flux of cocoon neutrinos could reach a comparable level as that of jet neutrinos, especially, being dominating at PeV energies. For a larger $E_c$, the photospheric radius is larger according to equation (5) and the proton cooling is less severe. This could result in a higher flux of PeV neutrinos. Most importantly, since the opening angle of the cocoon is larger $\theta_c\sim0.5$, its neutrinos have greater chance to be observed by the future IceCube-Gen2, though the correlation with GRBs may not be easy to identify since we need a successful observation of the cocoon's electromagnetic emission as a prerequisite.

\begin{figure}
\begin{center}
\plotone{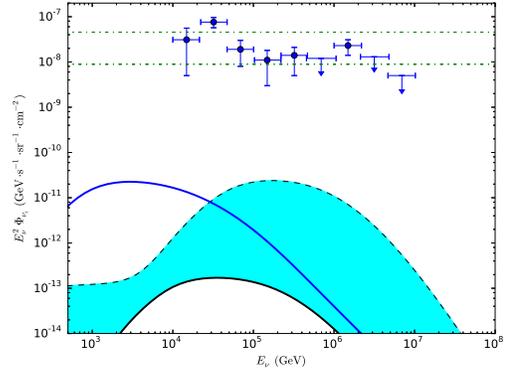}
\caption{The diffuse prompt neutrino flux (per flavor) of GRBs. The blue solid line represents the jet neutrinos and the diffuse flux that could be contributed by the cocoons is marked by the cyan region. The green dashed lines are the Waxman-Bahcall bound. Also shown is the diffuse neutrino background from IceCube observations (blue data-points).
\label{fig5}}
\end{center}
\end{figure}

\section{Discussions and Conclusions}
\label{sec:disc}

In this work we have revisited the dissipative photospheric scenario of GRBs, focusing on their neutrino emission. The dissipation mechanism is assumed to be internal shocks by which the CRs can be accelerated. The spectrum of a single GRB event is characterized by two suppression factors, which are caused by proton cooling and meson cooling respectively. The neutrino spectrum of the jet and the cocoon peak at different energies. Under conservative estimate, the neutrino flux from the cocoon is fainter than that from the jet, and needs longer waiting time for a successful observation. However, this depends on how much energy is deposited into the cocoon. If an GRB is by chance located at the Mpc distance, we have a greater chance to observe the cocoon neutrinos since the opening angle is larger than that of the jet. At last, we calculated the diffuse prompt neutrino flux for GRBs and found that the contribution from the cocoon could reach a comparable level as that of the jet. Note that since the diffuse flux depends on the rate and luminosity function of GRBs, we illustrate this effect in Figure 6, in which we involve two additional cases. One uses the cosmic GRB rate given by \citet{lie14} that has the same form as \citet{wand10} but with different best-fitting parameters (i.e., $z_1=3.6,\,n_1=2.07,\,n_2=-0.70,\,L_b=10^{52.05}\,\rm erg\,s^{-1},\,\alpha=0.65$, and $\beta=3.0$) (shown by the green solid line in Figure 6). The result of \citet{lie14} suggests the possibility of an intrinsic GRB rate that contains more bursts at higher redshifts than \citet{wand10} but the number of low-luminosity GRBs is suppressed. In this case, the diffuse neutrino flux is slightly higher. The other case uses the latest statistic burst rate that indicates an excess at low redshifts and for which the luminosity evolution is involved \citep{yu15, pet15} (shown by the red solid line in Figure 6), thus leading to a much higher diffuse neutrino flux.

\begin{figure}
\begin{center}
\plotone{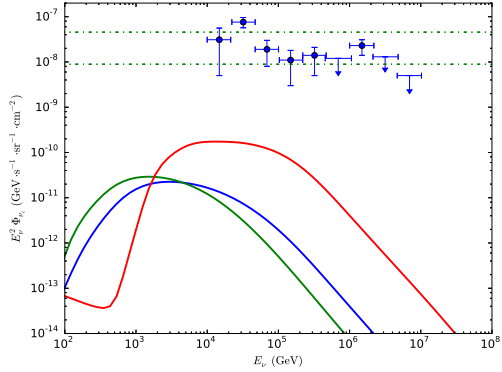}
\caption{The diffuse prompt neutrino flux (per flavor) of GRBs when we adopt different cosmic GRB rates and luminosity functions and take jet neutrinos as an example (note that the trend is the same for cocoon neutrinos). The blue solid line represents the diffuse flux of jet neutrinos when adopting the burst rate given by \citet{wand10}, and similarly, the green line corresponds to \citet{lie14} and the red line to \citet{yu15}.
\label{fig6}}
\end{center}
\end{figure}

IceCube reported the detection of five low-significance events correlated with five GRBs. However, these events are consistent with the background expectation from atmospheric muons and neutrinos \citep{aar16}. Lately, extended search for neutrinos coincident with GRBs in IceCube data shows no significant correlation \citep{aar17}. On one hand, the diffuse neutrino flux of GRB jets in the expectation of our model is less than one tenth of the IceCube observed diffuse background. On the other hand, the possibility of verifying the correlation is sufficiently cut down since the jet is highly beamed. One of every $\sim 400$ GRBs has a jet pointing to us. Since the cocoon has a larger opening angle, if the observing angle is in the range of $\theta_j<\theta_{\rm obs}<\theta_c$, the only neutrino signal we can observe is from the cocoon. However, since we would miss the GRB prompt emission in this situation, this neutrino signal from the GRB cocoon still could not verify the correlation of a GRB and a HESE, although physically speaking they should be correlated. However, with the accumulating data of HESE and all-sky transient survey, there is still a good chance to test the correlation in the near future.

\acknowledgements
We thank the referee for valuable comments and constructive suggestions. This work was supported by the National Basic Research Program of China (973 Program grant 2014CB845800) and the National Natural Science Foundation of China grant 11573014 (D.X. and Z.G.D.) and by NASA NNX13AH50G (P.M.)

\clearpage

\end{document}